\documentclass[onecolumn, manuscript]{revtex4}
\usepackage{graphicx}
\begin{document}

\title{Spiraling of adjacent trajectories in chaotic systems}
\author{P.V. Elyutin}
\email{pve@shg.phys.msu.su} \affiliation {Department of Physics,
Moscow State University, Moscow 119992, Russia}
\date{21 June 2004}

\begin{abstract}
The spiraling of adjacent trajectories in chaotic dynamical
systems can be characterized by the distribution of local angular
velocities of rotation of the displacement vector, which is
governed by linearized equations of motion. This distribution,
akin to that of local Lyapunov exponents, is studied for three
examples of three-dimensional flows. Toy model shows that the
rotation rate of adjacent trajectories influences on the rate of
mixing of dynamic variables and on the sensitivity of trajectories
to perturbations. \vspace{5mm}

 \vspace{5mm}
\end{abstract}
\maketitle

\vspace{10mm}
\section{Introduction} In nonlinear dynamics of systems - flows
with equations of motion
\begin{equation}\label{1}
\dot {\mathbf x}={\mathbf F} \left( { {\mathbf x}} \right),
\end{equation}
where ${\mathbf x}(t)$ is a vector representing the state of the
system in a $K$-dimensional phase space, ${\mathbf x} ={\left\{
{x_i} \right\}}$, $1 \le i \le K$, definitive role is played by
the properties of evolution of the displacement vector ${\mathbf
r} (t)$ that is governed by the linear system of equations
\begin{equation}\label{2}
\dot{ \mathbf r} =\hat M {\mathbf r}.
\end{equation}
Here $\hat M$ is the local stability matrix with the elements
$M_{ij}={{\partial F_i} \mathord{\left/ {\vphantom {{\partial
F_i} {\partial x_j}}} \right. \kern-\nulldelimiterspace}
{\partial x_j}}$, that are taken at the points of the phase
trajectory ${\mathbf x}(t)$ and depend, generally, on time. The
existence of the exponentially growing solutions of the system (2)
serves as a definition of chaoticity of motion of the system (1).

For almost any initial conditions the displacement vector
${\mathbf r} (t)$ will approach the eigenmode of the system (2)
that has the largest rate of the exponential growth, given by the
first (maximal) characteristic Lyapunov exponent $\sigma_1$ or
simply the Lyapunov exponent $\sigma$ \cite{LL92,MP00}.   Thus the
behaviour of  ${\mathbf r} (t)$ at large times $t \gg \sigma^{-1}$
loses its dependence on the initial conditions and represents some
characteristics of the chaotic component as a whole.  This
property serves as a base for the standard method of numerical
calculation of the Lyapunov exponent \cite{BGS76}, that is
determined from the rate of growth of the length of vector
${\mathbf r} (t)$. However, the evolution of the orientation of
this vector is usually ignored.

We restrict ourselves by the phase space with dimensionality
$K=3$, the minimal one in which the chaotic motion of systems -
flows like (1) is possible. Then the evolution of the direction of
the displacement vector ${\mathbf r} (t)$ can be described as a
rotation around the direction of the phase trajectory: the
adjacent trajectories can be spiraling, as it is shown in figure
1.
\newpage

\begin{figure}
[!ht]
\includegraphics[width=0.4\columnwidth]{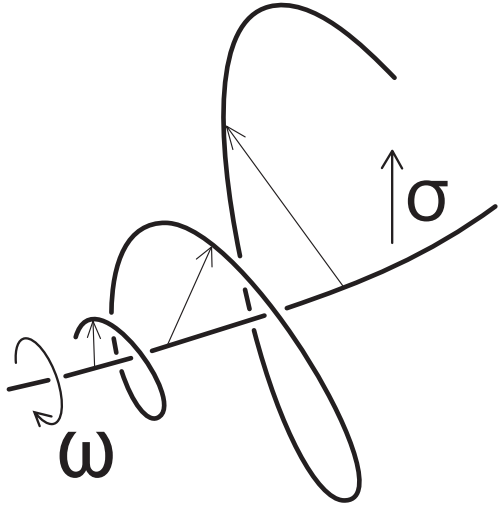}
\caption{\label{fig1} Qualitative scheme of position of adjacent
phase trajectories for chaotic motion in three-dimensional phase
space. Whereas the displacement vector grows in length by the
exponential law with the rate $\sigma$, it may also rotate with
the angular velocity $\omega$}
\end{figure}

The instantaneous angular velocity of this rotation $\omega(t)$
(we shall also call it "the rotation rate") is defined by the
expression
\begin{equation}\label{3}
\omega (t) = \lim_{\epsilon \to 0} {{1}\over{\epsilon}} \cdot
{{{\mathbf r}(t+\epsilon)\times{\mathbf r} (t)\cdot {\mathbf
v}(t)} \over{\|{\mathbf r}(t+\epsilon)\|\|{\mathbf
r}(t)\|\|{\mathbf v}(t)\|}},
\end{equation}
where ${\mathbf v}(t)=\dot{\mathbf x}(t)$ is a vector of phase
velocity, that is tangent to the phase trajectory, and ${\mathbf
a}\times {\mathbf b}\cdot {\mathbf c}$ denotes the mixed product
of three vectors.

The distribution $W$ of values of $\omega(t)$ does not depend on
the choice of the initial moment of time or the initial conditions
for the system (2) as far as they belong to the same chaotic
component. It is as universal as the distribution of the local
Lyapunov exponents that are defined by the expression
\begin{equation}\label{4}
\sigma (t) = \lim_{\epsilon \to 0} {{1}\over{\epsilon}} \ln
{{\|{\mathbf r}(t+\epsilon)\|}\over{\|{\mathbf r}(t)\|}}.
\end{equation}
The latter quantity is widely used for studies of the structure of
the phase space of chaotic systems \cite{SBP89,AB93,PR99}.  From
comparison of equations (\ref{3}) and (\ref{4}) one can assume
that the instantaneous angular velocity in Pickwickian sense can
be considered as an imaginary part of the local Lyapunov exponent
and could be used as a probe for exploration of fine details of
chaotic dynamics.

\section{Three examples}
The distribution $W(\omega)$ could be easily obtained numerically.
As the first example we consider the famous Lorenz model
\cite{L63} given by the equations of motion
\begin{eqnarray}\label{5}
\nonumber\dot X&=&-\sigma X+\sigma Y,\\
\dot Y &=&-XZ+rX-Y,\\
\nonumber \dot Z&=& XY-bZ,
\end{eqnarray}
with the standard values of parameters $\sigma = 10$, $r=28$,
$b=8/3$.  The distribution $W(\omega)$ is shown in figure 2.

\begin{figure}
[!ht]
\includegraphics[width=0.6\columnwidth]{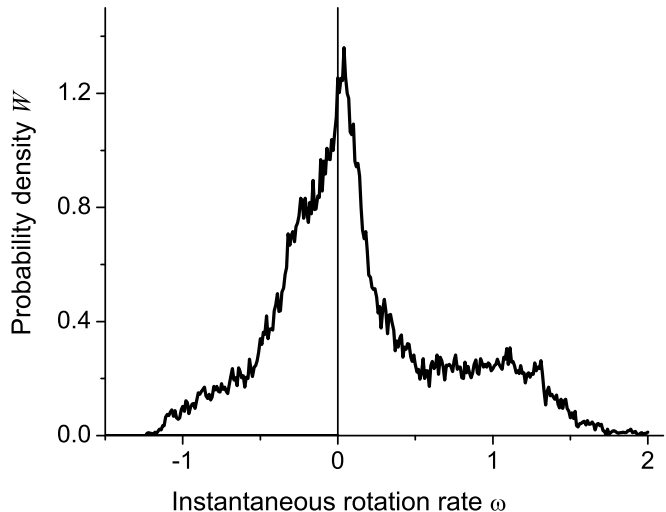}
\caption{\label{fig2} Distribution $W$ of the instantaneous
rotation rate $\omega$ for the Lorenz model defined by the system
of equations (5) with the standard values of parameters $\sigma =
10$, $r=28$, $b=8/3$.}
\end{figure}

The average value of the rate of rotation $\overline {\mathstrut
\omega}  = 0.158$ is much less than the width of the distribution
(standard deviation of $\omega$) $\Delta \omega = 0.63$. It may be
noted that for the Lorenz model with the used values of parameters
the characteristic frequency of motion $\Omega$ estimated from the
power spectrum of the variable $Z(t)$ is about $\Omega \approx 8$,
whereas the Lyapunov exponent $\sigma \approx 1.0$.  Thus the
small value of $\overline {\mathstrut \omega}$ may indicate the
presence of the additional time scale of motion.  We also note
that the time of decay of correlations of $\omega (t)$ is rather
close to $\tau \approx \Omega^{-1}$.

For the second example we take a conservative autonomous system,
the Pullen - Edmonds model oscillator \cite{PE81} with the
Hamiltonian
\begin{equation} \label{6}
H={{\mathbf p}^2 \over 2}+U(x,y)={1 \over 2}\left( {p_x^2+p_y^2}
\right)+{1 \over 2}\left( {x^2+y^2+x^2y^2} \right),
\end{equation}
where $x,y$ are the Cartesian coordinates of the particle and
$p_x,p_y$ are the corresponding canonically conjugated momenta. In
expression (6) the particle mass $m$, the frequency of small
oscillations $\omega_0$ and the nonlinearity length $\lambda$ have
been used as the unit scales. The motion of this system takes
place on a three-dimensional energy surface in the
four-dimensional phase space. Since the energy of the system is
conserved, $E\equiv H=\mathrm{const}$, the absolute value of the
total momentum $ p\left( {x,y} \right)=\sqrt {2\left[ {E-U\left(
{x,y} \right)} \right]}$ is completely defined by the position of
the phase point in the configurational space. The angle $\varphi$
that gives the direction of the total momentum is defined by the
equations
\begin{equation} \label{7}
p_x=p\,\cos \varphi ,\,\,\,\,\,p_y=p\,\sin \varphi.
\end{equation}
It can be taken as the third dynamic variable, that yields the
equations of motion
\begin{equation} \label{8}
\dot x=p\cos \varphi ,\,\,\,\,\,\,\, \dot y=p\sin \varphi
,\,\,\,\,\,\, \dot \varphi ={1 \over p}\left( {{{\partial U} \over
{\partial x}}\sin \varphi -{{\partial U} \over {\partial y}}\cos
\varphi } \right).
\end{equation}
For the motion on the energy surface $E=15$, that is nearly
ergodic \cite{M86}, the distribution $W(\omega)$ is shown in
figure 3.

\begin{figure}
[!ht]
\includegraphics[width=0.6\columnwidth]{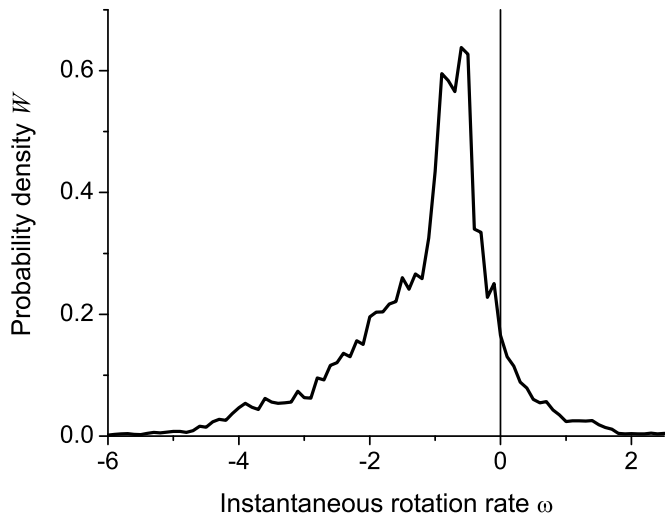}
\caption{\label{fig3} Distribution $W$ of the instantaneous
rotation rate $\omega$ for the Pullen - Edmonds model, defined by
the system of equations (8), on the energy surface $E-15$.}
\end{figure}

The average value of the rate of rotation $\overline {\mathstrut
\omega} = -1.49$ is comparable to the width of the distribution
$\Delta \omega = 1.53$ and to the characteristic frequency of
motion $\Omega \sim 1$.

For the third example we take a conservative non-autonomous system
with the equation of motion
\begin{equation} \label{9}
\ddot x+x^3=F \cos \Omega t,
\end{equation}
that describes an oscillator with cubic nonlinearity under the
influence of external harmonic force (Duffing model). The system
(9) could be turned into an autonomous three-dimensional system by
introducing dynamical variables $x$, $y \equiv \dot x$ and
interpreting the time $t$ in the RHS as a third dynamic variable
$z$ that obeys the equation of motion $\dot z=1$ \cite{LL92}. For
the motion in the chaotic component that surrounds the line $x=0$,
$y=0$ the distribution $W(\omega)$ is shown in figure 4.

\begin{figure}
[!ht]
\includegraphics[width=0.6\columnwidth]{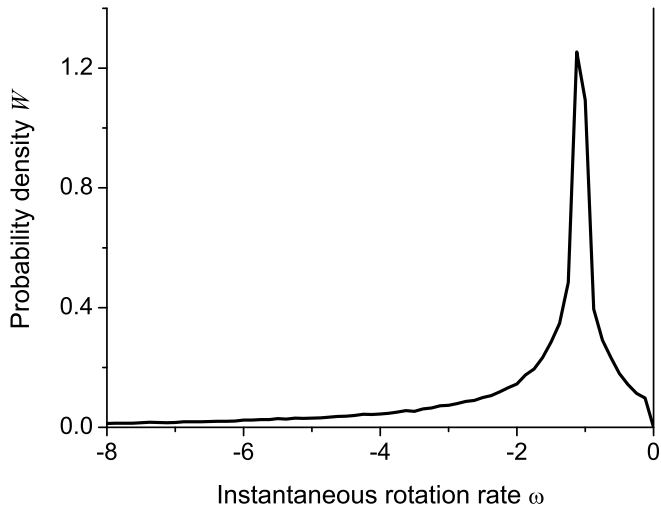}
\caption{\label{fig4} Distribution $W$ of the instantaneous
rotation rate $\omega$ for the Duffing model defined by equation
(9) with parameters $F=5$, $\Omega=1$.}
\end{figure}

It follows from equation (9) that the rotation rate is always
negative.  Its average value $\overline {\mathstrut \omega} =
-2.24$ is comparable to the width of the distribution $\Delta
\omega = 2.38$.  We note that the value of $\overline {\mathstrut
\omega}$ is not trivial: it is incommensurate with the driving
frequency $\Omega = 1$.

From comparison of graphs in figures 2 - 4 it may be noticed that
although our examples were taken from very different classes of
dynamical systems, the forms of distributions have some common
features: they are asymmetric, strongly peaked at a single
non-zero most probable value and have extended wings.

\vspace{5mm}

\section{Spiraling and mixing}
The proof of usefulness of the quantity $\omega (t)$ for the
exploration of the phase space we leave for the future studies.
There is another way to reveal the relevance of $\omega(t)$ to the
properties of chaotic motion: to find a relation between this
quantity and some established characteristics of chaoticity. The
affinity between $\overline {\mathstrut \omega}$ and $\sigma$
suggests that relations that include the Lyapunov exponent may be
extended to include $\omega$ in a similar way.

In nonlinear dynamics there is a widely known rule of thumb
stating that the rate of mixing (the inverse time of damping of
correlations of dynamic variables) $\gamma$ is approximately
equal to the Lyapunov exponent:
\begin{equation} \label{10}
\gamma \approx \sigma.
\end{equation}
This relation sometimes turns into exact equality (\textit{e.g.}
for the linear random number generator $x' = \{ Qx \}$, where
$\{,\}$ denotes the fractional part of a number, with integer
parameter $Q$); sometimes equation (10) reflects the scaling
property $\gamma \propto \sigma$ (\textit{e.g.} it holds exactly
for billiards), and in general it is expected to be fulfilled at
least semiquantitatively for arbitrary systems \cite{Z84}. From
the picture in figure 1 one could surmise that the increase of
$\omega$ will intensify the rate of exploration of the phase space
and thus enhance the mixing.

For the check of this hypothesis we retreat to two-dimensional
mappings, that can be interpreted as Poincare mappings for
three-dimensional flows.  In this class it is easy to construct a
model that permits controllable and independent changes of
$\omega$ and $\sigma$. Let $\{\xi_n,\eta_n\}$ be the components of
the vector ${\mathbf s}_n$ of normalized displacement at the
$n$-th iteration, ${\mathbf s}_n={\mathbf r}_n /\|{\mathbf
r}_n\|$. Then the rate of rotation is equal to the angle of
rotation for one step:
\begin{equation}\label{11}
\omega_{n+1}=\arcsin \left({\xi_{n+1} \eta_n-\xi_n
\eta_{n+1}}\right).
\end{equation}

Let's define the rate of damping of correlations (rate of mixing)
of a dynamical variable with the known correlation function $B
(n)$ by the relation

\begin{equation}\label{12}
\gamma =\ln \left( {{S \over {S-1}}} \right),\,\,\,\,\,\,S={1
\over {B\left( 0 \right)}}\sum\limits_{i=0}^\infty  {\left|
{B\left( n \right)} \right|}
\end{equation}
For the exponentially decreasing correlation function $B\left( n
\right)=B\left( 0 \right)\exp \left( {-\alpha n} \right)$ this
definition produces the rate of mixing equal to the exponent,
$\gamma = \alpha$.

For the study of influence of rotation on mixing we take a
two-dimensional mapping for which it is possible to variate
rotation rate $\omega$ with constant $\sigma$, namely a piecewise
linear mapping of a unit square on itself
\begin{equation}\label{13}
x'=\left\{ {ax+by} \right\},\,\,\,\,y'=\left\{ {-bx+ay} \right\},
\end{equation}
where
\begin{equation}\label{14}
a=\exp \sigma \cos \omega ,\,\,\,\,\,\,\,b=\exp \sigma \sin
\omega.
\end{equation}
The behaviour of rates of mixing for dynamical variables $x$ and
$y$ is shown in figure 5. \vspace{10mm}

\begin{figure}
[!ht]
\includegraphics[width=0.7\columnwidth]{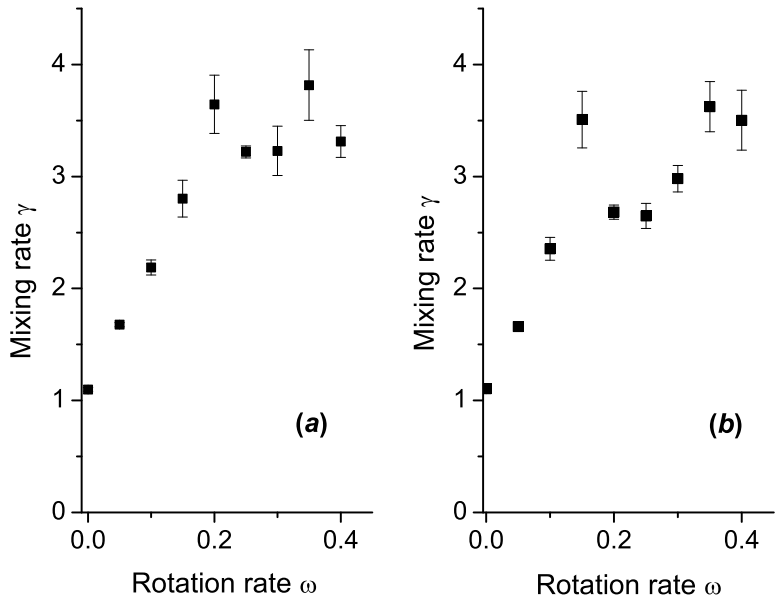}
\caption{\label{fig5} Dependence of the mixing rate $\gamma$ on
the rotation rate $\omega$ for the two-dimensional mapping (13):
(\textit{a}) - for variable $x$; (\textit{b}) - for variable $y$.}
\end{figure}

The results of numerical calculation confirms the intuitive guess:
at small values of $\omega$ its increase leads to the increase of
the rate of mixing $\gamma$. \vspace{15mm}

\section{Spiraling and response to perturbations}
Let's compare the system (1) to its perturbed version with the
equations of motion
\begin{equation}\label{15}
\dot { \textbf{x}}= \textbf{F}\left( {\textbf{x}}
\right)+\varepsilon  \textbf{V}\left( {\textbf{x}} \right).
\end{equation}
where $\varepsilon \textbf{V}\left( {\textbf{x}}\right )$ is some
static perturbation.  Let $\textbf{x}_0(t)$ and $\textbf{x}_V(t)$
be the laws of motion of the unperturbed system (1) and perturbed
system (15) correspondingly with the same initial conditions, and
let $\Delta \left( t \right)=\left| {\textbf{x}_0\left( t
\right)-\textbf{x}_V\left( t \right)} \right|$ denote the distance
between the phase trajectories at a given moment $t$. For a given
point of the phase space let's define a quantity
\begin{equation}\label{16}
\Delta \left( {\textbf{x}} \right)=\mathop {\lim
}\limits_{\varepsilon \to 0,t\to \infty }{{\Delta \left( t
\right)} \over {\left| \varepsilon  \right|}}\exp \left( {-\sigma
t} \right),
\end{equation}
that defines the amplitude of the response of the system to the
perturbation of a given form.  This quantity, averaged over a
given chaotic component of the phase space, $\overline{\mathstrut
\Delta}
 =\left\langle {\Delta \left( {\textbf{x}}
\right)} \right\rangle$, gives a convenient measure of sensitivity
of the system (1) to small perturbations.

Let's treat the simplest model of the evolution of deviations
given by a system of flows
\begin{equation}\label{17}
\dot \xi -\sigma \xi +\omega \eta =U,\,\,\,\,\,\,\dot \eta -\omega
\xi -\sigma \eta =V,
\end{equation}
with constant parameters $\sigma$, $\omega$, $U$ and $V$.  In the
absence of perturbations ($U=V=0$) it describes the rotation of
the displacement vector with the constant angular velocity
$\omega$ and its exponential growth with the exponent $\sigma$.
For $U,V \neq 0$ the asymptotics of solutions for $\sigma t \gg 1$
have the form
\begin{eqnarray}\label{18}\nonumber
 \xi \approx {{e^{\sigma t}} \over {\sigma ^2+\omega ^2}}\left[
{\left( {\sigma U-\omega V} \right) \cos \omega t+\left( {\omega
U+\sigma V} \right)\sin \omega t} \right],\\
\eta \approx {{e^{\sigma t}} \over {\sigma ^2+\omega ^2}}\left[
{\left( {\omega U+\sigma V} \right) \cos \omega t+\left( {-\sigma
U+\omega V} \right)\sin \omega t} \right].
\end{eqnarray}
By definition (16) for each of the perturbations we have
\begin{equation}\label{19}
\overline {\mathstrut \Delta} ={1 \over {\sqrt {2\left( {\sigma
^2+\omega ^2} \right)}}}.
\end{equation}
Therefore the schematic model (17) shows that with the increase of
the rotation rate the amplitude of the response will decrease. One
may infer that this connection may be universal.

Let's return to the piecewise model (13) perturbed by small
variations of its control parameters:
\begin{equation}\label{20}
x'=\left\{ {\left( {a+\varepsilon } \right)x+by}
\right\},\,\,\,\,y'=\left\{ {-bx+ay} \right\}\,\,\,
\end{equation}
and
\begin{equation}\label{21}
x'=\left\{ {ax+\left( {b+\varepsilon } \right)y}
\right\},\,\,\,\,y'=\left\{ {-bx+ay} \right\}.\,\,\,
\end{equation}

Fig. 6 demonstrates that with the increase of the rotation rate
$\omega$ the averaged amplitude of response to the perturbation
$\overline{\mathstrut \Delta}$ in general decreases, albeit
non-monotonously, in analogy with the dependence (19).

\begin{figure}
[!ht]
\includegraphics[width=0.7\columnwidth]{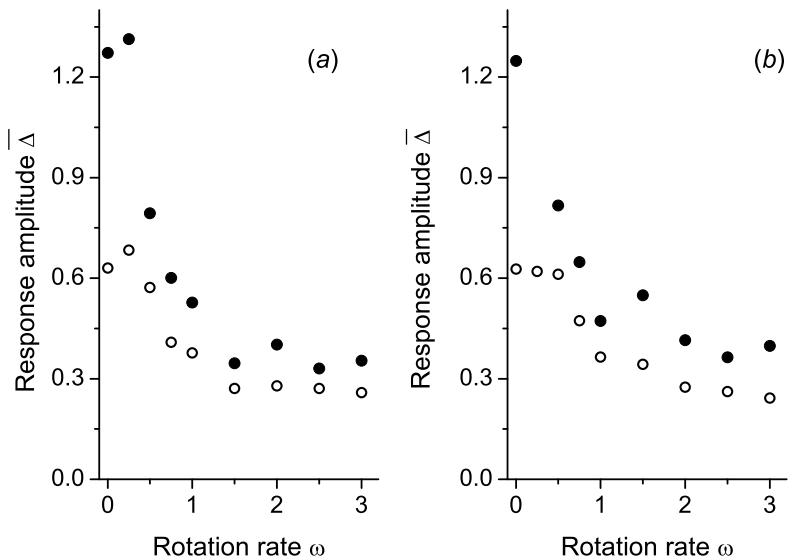}
\caption{\label{fig6} Dependence of the response amplitude
$\overline {\Delta}$ on the rotation rate $\omega$ for different
perturbations of the two-dimensional mapping (13): (\textit{a}) -
perturbed system is given by equation (20); (\textit{b}) -
perturbed system is given by equation (21). In both pictures
filled circles are for $\sigma = 0.25$ and open circles - for
$\sigma = 0.5$.}
\end{figure}

Recently the changes of the evolution under the influence of small
perturbation have attracted considerable attention in the theory
of Hamiltonian chaotic systems in connection with the problem of
dynamics of fidelity \cite{JP01, BC02, KJZ02, JAB02}.  The example
described in this section shows that the rotation rate can display
itself in the preexponential factors of this response.\vspace{5mm}

\section{Generalization to higher dimensions}
Everywhere above we have restricted ourselves to analysis of
trajectories in three-dimensional phase space.  The generalization
of the definition (3) to higher dimensions could be obtained
straightforwardly from the expression for the oriented volume
$V_K$ in $K$-dimensional space \cite{PSh90}.  It can be
constructed from the unit vector along the direction of phase
velocity ${\mathbf u}(t)={\mathbf v}(t)/\|{\mathbf v}(t)\|$ and
($K-1$) unit vectors along the displacement vectors taken in
consequent equidistant moments of time, ${\mathbf
s}(t+k\epsilon)={\mathbf r}(t+k\epsilon) /\|{\mathbf
r}(t+k\epsilon)\|$, $1 \leq k \leq K-2$ by their convolution with
the completely antisymmetric tensor of the $K$-th rank
$\textsf{E}_{ijk...n}$:

\begin{equation}\label{22}
V_K\left(t \right)=\textsf{E}_{ijk...n} u_i s_j\left( t \right)
s_k\left( {t+\epsilon} \right)....s_n\left( {t+\left({K-2}
\right)\epsilon} \right).
\end{equation}
However, this quantity scales as $\epsilon^N$, where
$N=(K-1)(K-2)/2$, and the generalized rate of "rotation",
\begin{equation}\label{23}
\omega_K(t)=\lim_{\epsilon \to 0}\epsilon^{-N}V_K(t),
\end{equation}
for $K>3$ depends not only on the elements of stability matrix at
a given moment of time but also on their time derivatives up to
the order $N-1$.\vspace{5mm}

\section*{Acknowledgements}
The author appreciate valuable discussions with E.D. Belega, L.V.
Keldysh, G.N. Medvedev, A.A. Nikulin, and D.D. Sokolov. The author
is especially grateful to A.A. Nikulin for his artistic drawing of
figure 1.  The author acknowledges the support by the "Russian
Scientific Schools" program (grant \# NSh - 1909.2003.2).

\end{document}